\documentclass[aps,prl,preprint,tightenlines,superscriptaddress,showpacs,byrevtex]{revtex4}
\usepackage{afterpage,rotating}
\usepackage{currvita}

\usepackage{tabularx}
\usepackage{nonfloat}
\usepackage{graphicx}
\usepackage{dcolumn}
\usepackage{bm}
\usepackage{subfigure}
\usepackage{mathrsfs}
\usepackage{amsmath,amsbsy,amssymb,lmodern}
\usepackage{epsfig,color}
\usepackage{placeins}
\usepackage{rccol}
\usepackage{comment}
\usepackage{hyperref} 
\usepackage{setspace} 
\usepackage{tabularx} 
\usepackage{multirow}
\usepackage{braket}
\usepackage{listings}

\usepackage{floatrow}
\usepackage{float}

\usepackage{accents}

\usepackage[letterpaper,portrait,lmargin=1.0in,rmargin=1.0in,tmargin=1.0in,bmargin=1.0in]{geometry}

\newcommand{\bseep}{B_s^0 \rightarrow \eta^\prime \eta}

\newcommand{\bksep}{B^0 \rightarrow \eta^\prime K_S^0}

\newcommand\mybar[1]{\accentset{\rule{0.6em}{0.6pt}}{#1}}

\newcommand{\mbc}{M_{\rm bc} }

\newcommand{\de}{\Delta E}

\newcommand{\mep}{M(\pi^+\pi^-\eta)}

\newcommand{\beep}{B_s^0 \rightarrow \eta^\prime \eta}
\newcommand{\et}{\eta}
\newcommand{\bst}{B_s^{*0}}
\newcommand{\bs}{B_s^0}
\newcommand{\fbssbss}{f_{B_{s}^{*0} \mybar{B}_s^{*0}}}
\newcommand{\fbsbss}{f_{B_{s}^{*0} \mybar{B}_s^0}}
\newcommand{\etp}{\eta^{\prime}}

\newcommand{\bsbs}{B_s^{*0} \mybar{B}_s^{*0}$, $B_s^{*0} \mybar{B}_s^0$ or $B_s^{0} \mybar{B}_s^{*0}$, and $ B_s^0 \mybar{B}_s^0}

\newcommand{\bsbsbara}{B_s^{(*)0}\mybar{B}_s^{(*)0}}

\begin{document}


\preprint{\vbox{ \hbox{   }
                 \hbox{BELLE-CONF-2002}
}}


\title{ \quad\\[1.0cm] Search for the Decay $\beep$}

\noaffiliation
\affiliation{University of the Basque Country UPV/EHU, 48080 Bilbao}
\affiliation{Beihang University, Beijing 100191}
\affiliation{University of Bonn, 53115 Bonn}
\affiliation{Brookhaven National Laboratory, Upton, New York 11973}
\affiliation{Budker Institute of Nuclear Physics SB RAS, Novosibirsk 630090}
\affiliation{Faculty of Mathematics and Physics, Charles University, 121 16 Prague}
\affiliation{Chiba University, Chiba 263-8522}
\affiliation{Chonnam National University, Gwangju 61186}
\affiliation{University of Cincinnati, Cincinnati, Ohio 45221}
\affiliation{Deutsches Elektronen--Synchrotron, 22607 Hamburg}
\affiliation{Duke University, Durham, North Carolina 27708}
\affiliation{University of Florida, Gainesville, Florida 32611}
\affiliation{Department of Physics, Fu Jen Catholic University, Taipei 24205}
\affiliation{Key Laboratory of Nuclear Physics and Ion-beam Application (MOE) and Institute of Modern Physics, Fudan University, Shanghai 200443}
\affiliation{Justus-Liebig-Universit\"at Gie\ss{}en, 35392 Gie\ss{}en}
\affiliation{Gifu University, Gifu 501-1193}
\affiliation{II. Physikalisches Institut, Georg-August-Universit\"at G\"ottingen, 37073 G\"ottingen}
\affiliation{SOKENDAI (The Graduate University for Advanced Studies), Hayama 240-0193}
\affiliation{Gyeongsang National University, Jinju 52828}
\affiliation{Department of Physics and Institute of Natural Sciences, Hanyang University, Seoul 04763}
\affiliation{University of Hawaii, Honolulu, Hawaii 96822}
\affiliation{High Energy Accelerator Research Organization (KEK), Tsukuba 305-0801}
\affiliation{J-PARC Branch, KEK Theory Center, High Energy Accelerator Research Organization (KEK), Tsukuba 305-0801}
\affiliation{Higher School of Economics (HSE), Moscow 101000}
\affiliation{Forschungszentrum J\"{u}lich, 52425 J\"{u}lich}
\affiliation{Hiroshima Institute of Technology, Hiroshima 731-5193}
\affiliation{IKERBASQUE, Basque Foundation for Science, 48013 Bilbao}
\affiliation{University of Illinois at Urbana-Champaign, Urbana, Illinois 61801}
\affiliation{Indian Institute of Science Education and Research Mohali, SAS Nagar, 140306}
\affiliation{Indian Institute of Technology Bhubaneswar, Satya Nagar 751007}
\affiliation{Indian Institute of Technology Guwahati, Assam 781039}
\affiliation{Indian Institute of Technology Hyderabad, Telangana 502285}
\affiliation{Indian Institute of Technology Madras, Chennai 600036}
\affiliation{Indiana University, Bloomington, Indiana 47408}
\affiliation{Institute of High Energy Physics, Chinese Academy of Sciences, Beijing 100049}
\affiliation{Institute of High Energy Physics, Vienna 1050}
\affiliation{Institute for High Energy Physics, Protvino 142281}
\affiliation{Institute of Mathematical Sciences, Chennai 600113}
\affiliation{INFN - Sezione di Napoli, 80126 Napoli}
\affiliation{INFN - Sezione di Torino, 10125 Torino}
\affiliation{Advanced Science Research Center, Japan Atomic Energy Agency, Naka 319-1195}
\affiliation{J. Stefan Institute, 1000 Ljubljana}
\affiliation{Kanagawa University, Yokohama 221-8686}
\affiliation{Institut f\"ur Experimentelle Teilchenphysik, Karlsruher Institut f\"ur Technologie, 76131 Karlsruhe}
\affiliation{Kavli Institute for the Physics and Mathematics of the Universe (WPI), University of Tokyo, Kashiwa 277-8583}
\affiliation{Kennesaw State University, Kennesaw, Georgia 30144}
\affiliation{King Abdulaziz City for Science and Technology, Riyadh 11442}
\affiliation{Department of Physics, Faculty of Science, King Abdulaziz University, Jeddah 21589}
\affiliation{Kitasato University, Sagamihara 252-0373}
\affiliation{Korea Institute of Science and Technology Information, Daejeon 34141}
\affiliation{Korea University, Seoul 02841}
\affiliation{Kyoto Sangyo University, Kyoto 603-8555}
\affiliation{Kyoto University, Kyoto 606-8502}
\affiliation{Kyungpook National University, Daegu 41566}
\affiliation{Universit\'{e} Paris-Saclay, CNRS/IN2P3, IJCLab, 91405 Orsay}
\affiliation{\'Ecole Polytechnique F\'ed\'erale de Lausanne (EPFL), Lausanne 1015}
\affiliation{P.N. Lebedev Physical Institute of the Russian Academy of Sciences, Moscow 119991}
\affiliation{Liaoning Normal University, Dalian 116029}
\affiliation{Faculty of Mathematics and Physics, University of Ljubljana, 1000 Ljubljana}
\affiliation{Ludwig Maximilians University, 80539 Munich}
\affiliation{Luther College, Decorah, Iowa 52101}
\affiliation{Malaviya National Institute of Technology Jaipur, Jaipur 302017}
\affiliation{University of Malaya, 50603 Kuala Lumpur}
\affiliation{University of Maribor, 2000 Maribor}
\affiliation{Max-Planck-Institut f\"ur Physik, 80805 M\"unchen}
\affiliation{School of Physics, University of Melbourne, Victoria 3010}
\affiliation{University of Mississippi, University, Mississippi 38677}
\affiliation{University of Miyazaki, Miyazaki 889-2192}
\affiliation{Moscow Physical Engineering Institute, Moscow 115409}
\affiliation{Graduate School of Science, Nagoya University, Nagoya 464-8602}
\affiliation{Kobayashi-Maskawa Institute, Nagoya University, Nagoya 464-8602}
\affiliation{Universit\`{a} di Napoli Federico II, 80126 Napoli}
\affiliation{Nara University of Education, Nara 630-8528}
\affiliation{Nara Women's University, Nara 630-8506}
\affiliation{National Central University, Chung-li 32054}
\affiliation{National United University, Miao Li 36003}
\affiliation{Department of Physics, National Taiwan University, Taipei 10617}
\affiliation{H. Niewodniczanski Institute of Nuclear Physics, Krakow 31-342}
\affiliation{Nippon Dental University, Niigata 951-8580}
\affiliation{Niigata University, Niigata 950-2181}
\affiliation{University of Nova Gorica, 5000 Nova Gorica}
\affiliation{Novosibirsk State University, Novosibirsk 630090}
\affiliation{Osaka City University, Osaka 558-8585}
\affiliation{Osaka University, Osaka 565-0871}
\affiliation{Pacific Northwest National Laboratory, Richland, Washington 99352}
\affiliation{Panjab University, Chandigarh 160014}
\affiliation{Peking University, Beijing 100871}
\affiliation{University of Pittsburgh, Pittsburgh, Pennsylvania 15260}
\affiliation{Punjab Agricultural University, Ludhiana 141004}
\affiliation{Research Center for Electron Photon Science, Tohoku University, Sendai 980-8578}
\affiliation{Research Center for Nuclear Physics, Osaka University, Osaka 567-0047}
\affiliation{Meson Science Laboratory, Cluster for Pioneering Research, RIKEN, Saitama 351-0198}
\affiliation{Theoretical Research Division, Nishina Center, RIKEN, Saitama 351-0198}
\affiliation{RIKEN BNL Research Center, Upton, New York 11973}
\affiliation{Saga University, Saga 840-8502}
\affiliation{Department of Modern Physics and State Key Laboratory of Particle Detection and Electronics, University of Science and Technology of China, Hefei 230026}
\affiliation{Seoul National University, Seoul 08826}
\affiliation{Showa Pharmaceutical University, Tokyo 194-8543}
\affiliation{Soochow University, Suzhou 215006}
\affiliation{Soongsil University, Seoul 06978}
\affiliation{University of South Carolina, Columbia, South Carolina 29208}
\affiliation{Stefan Meyer Institute for Subatomic Physics, Vienna 1090}
\affiliation{Sungkyunkwan University, Suwon 16419}
\affiliation{School of Physics, University of Sydney, New South Wales 2006}
\affiliation{Department of Physics, Faculty of Science, University of Tabuk, Tabuk 71451}
\affiliation{Tata Institute of Fundamental Research, Mumbai 400005}
\affiliation{Excellence Cluster Universe, Technische Universit\"at M\"unchen, 85748 Garching}
\affiliation{Department of Physics, Technische Universit\"at M\"unchen, 85748 Garching}
\affiliation{School of Physics and Astronomy, Tel Aviv University, Tel Aviv 69978}
\affiliation{Toho University, Funabashi 274-8510}
\affiliation{Department of Physics, Tohoku University, Sendai 980-8578}
\affiliation{Earthquake Research Institute, University of Tokyo, Tokyo 113-0032}
\affiliation{Department of Physics, University of Tokyo, Tokyo 113-0033}
\affiliation{Tokyo Institute of Technology, Tokyo 152-8550}
\affiliation{Tokyo Metropolitan University, Tokyo 192-0397}
\affiliation{Tokyo University of Agriculture and Technology, Tokyo 184-8588}
\affiliation{Utkal University, Bhubaneswar 751004}
\affiliation{Virginia Polytechnic Institute and State University, Blacksburg, Virginia 24061}
\affiliation{Wayne State University, Detroit, Michigan 48202}
\affiliation{Yamagata University, Yamagata 990-8560}
\affiliation{Yonsei University, Seoul 03722}
  \author{A.~Abdesselam}\affiliation{Department of Physics, Faculty of Science, University of Tabuk, Tabuk 71451} 
  \author{I.~Adachi}\affiliation{High Energy Accelerator Research Organization (KEK), Tsukuba 305-0801}\affiliation{SOKENDAI (The Graduate University for Advanced Studies), Hayama 240-0193} 
  \author{K.~Adamczyk}\affiliation{H. Niewodniczanski Institute of Nuclear Physics, Krakow 31-342} 
  \author{J.~K.~Ahn}\affiliation{Korea University, Seoul 02841} 
  \author{H.~Aihara}\affiliation{Department of Physics, University of Tokyo, Tokyo 113-0033} 
  \author{S.~Al~Said}\affiliation{Department of Physics, Faculty of Science, University of Tabuk, Tabuk 71451}\affiliation{Department of Physics, Faculty of Science, King Abdulaziz University, Jeddah 21589} 
  \author{K.~Arinstein}\affiliation{Budker Institute of Nuclear Physics SB RAS, Novosibirsk 630090}\affiliation{Novosibirsk State University, Novosibirsk 630090} 
  \author{Y.~Arita}\affiliation{Graduate School of Science, Nagoya University, Nagoya 464-8602} 
  \author{D.~M.~Asner}\affiliation{Brookhaven National Laboratory, Upton, New York 11973} 
  \author{H.~Atmacan}\affiliation{University of Cincinnati, Cincinnati, Ohio 45221} 
% ...
\collaboration{The Belle Collaboration}


\begin{abstract}

\par We report a search for the decay $\beep$ using $121.4$ fb$^{-1}$ of data
collected at the $\Upsilon$(5S) resonance
with the Belle detector
at the KEKB asymmetric-energy electron-positron collider.
This decay is suppressed in the Standard Model 
and proceeds through transitions sensitive to new physics.
The expected branching fraction for $\beep$ in the Standard Model 
spans a wide range [$(2 - 4)\times10^{-5}$] 
with a large theoretical uncertainty 
due to non-perturbative hadronic physics.
We apply a discovery-optimized background suppression method 
and report a 90\% confidence-level upper limit of $7.1 \times 10^{-5}$ 
on the branching fraction for this decay. 

\end{abstract}

\pacs{XX.YY.ZZ, AA.BB.CC}

\maketitle

\tighten

{\renewcommand{\thefootnote}{\fnsymbol{footnote}}}
\setcounter{footnote}{0}

\section{Introduction and Physics motivation}

\begin{figure}[htb!]
\begin{minipage}[c]{0.33\linewidth}
\small
 \begin{center}
    \subfigure{\includegraphics[width=1.0\textwidth]{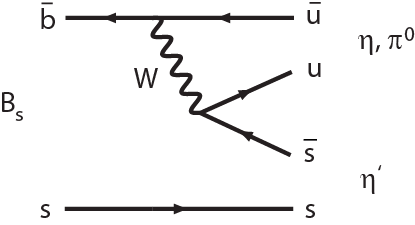}}
 \end{center}
\end{minipage}\hfill
\begin{minipage}[c]{0.33\linewidth}
\small
 \begin{center}
    \subfigure{\includegraphics[width=1.0\textwidth]{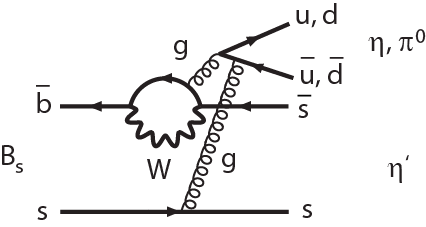}}
 \end{center}
\end{minipage}\hfill
\begin{minipage}[c]{0.27\linewidth}
\small
 \begin{center}
    \subfigure{\includegraphics[width=1.0\textwidth]{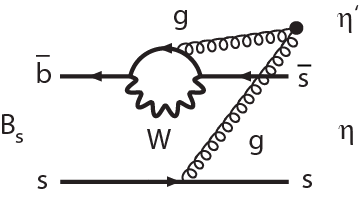}}
 \end{center}
\end{minipage}\hfill
\figcaption{ Tree level, gluonic penguin, and $\eta^\prime$ gluon admixture Feynman diagrams for
charmless two-body decays of $B_s^0$ to pairs of pseudoscalar mesons.
}
\label{fynmn}
\end{figure}

In the Standard Model (SM) charmless hadronic decays $\bseep$
proceed via tree-level $b\to u$ and penguin $b\to s$ transitions as shown in Fig.~\ref{fynmn}.
Penguin transitions are sensitive to  Beyond-the-Standard-Model (BSM) physics scenarios
and could affect the branching fractions and {\it CP} asymmetries in such decays\cite{belleiiphysicsbook}.
Once branching fractions for two-body decays $B_s \to \et\et, \et\etp, \etp\etp $ are measured,
and the theoretical uncertainties are reduced,
it would be possible to extract {\it CP} violating parameters
from the data using the formalism based on SU(3)/U(3) symmetry~\cite{bf1}.
To achieve this goal, at least four of these six branching fractions need to be measured.
Only the branching fraction for $B_s^0 \to \eta^{\prime}\eta^{\prime}$ has been measured so far~\cite{bsepep}.

\section{Data Sample and Belle Detector}

In this paper we report the results of the first search for the decay $\bseep$
using the full Belle data sample of $121.4\textrm{fb}^{-1}$
collected at the $\Upsilon(5S)$ resonance.
The Belle detector~\cite{Belle}
was 
a large-solid-angle magnetic spectrometer
that operated at the KEKB asymmetric-energy $e^+e^-$ collider~\cite{KEKB}.
The detector components relevant to our study include
a tracking system comprising a silicon vertex detector (SVD) and a central drift chamber (CDC),
a particle identification (PID) system
that consists of a barrel-like arrangement of time-of-flight scintillation counters (TOF)
and an array of aerogel threshold Cherenkov counters (ACC),
and a CsI(Tl) crystal-based electromagnetic calorimeter (ECL).
All these components are located inside a superconducting
solenoid coil that provides a 1.5~T magnetic field.

The $\Upsilon(5S)$ decays into
$\bsbs$ pairs
with relative fractions $\fbssbss =(87.0\pm1.7)\%$ and $\fbsbss=(7.3\pm1.4)\%$~\cite{frac}.
The data sample contains $(6.53\pm0.66)\times10^{6}$ $\bsbsbara$ pairs~\cite{nbsbsb}.
The excited vector state $\bst$ decays to $\bs$ by emitting a photon.
The daughter $\etp$ meson is reconstructed in the decay mode $\pi^+\pi^- \eta$, 
each of the two $\eta$ mesons is reconstructed via its two photon decay.
The expected branching fraction for the $B_s$ decay of interest spans 
a wide range: $(2 - 4)\times10^{-5}$~\cite{bf1, bf2, bf3, bf4, bf5}, where the main
source of theoretical uncertainty is due to non-perturbative hadronic physics.

To maximize analysis discovery potential 
and
to validate the signal extraction procedure
we use a background Monte Carlo (MC) sample equivalent to six times the data statistics.
We use a high-statistics signal MC sample to estimate the overall reconstruction efficiency.
Both samples are used to develop 
a model implemented 
in the unbinned extended maximum likelihood (ML) fit to data.
The MC-based model is calibrated using a control data sample 
of 711~${\rm fb^{-1}}$ collected at the $\Upsilon(4S)$. 

\section{Reconstruction and Signal Candidate Selection}

We reconstruct $\eta$ candidates
using pairs of electromagnetic showers
not matched to the projections
of charged tracks to the calorimeter.
We require that the reconstructed energy of these showers
exceed 50 (100) MeV in the barrel (end-cap)
region of the ECL. 
The larger end-cap ECL energy threshold 
is due to the larger beam-related background in this region. 
The ECL energy thresholds have practically 
no impact on the analysis discussed in this paper. 
To reject hadronic showers mimicking photons,
the ratio of the energy deposited by a photon candidate
in the $(3\times3)$ and $(5\times 5)$ ECL crystal array
centered on the crystal with the largest reconstructed energy
is required to exceed 0.75.
The invariant mass of the $\eta$ candidate is required
to be in the range $515 \le M(\gamma\gamma) \le 580$~${\rm MeV/c}^2$, 
which corresponds, approximately, to $\pm 3 \sigma$ 
when approximated by a Gaussian resolution function.
To suppress misreconstructed $\eta$ candidates,
the absolute value of cosine of helicity angle
(defined as the angle between
the photon momentum in presumed parent's rest frame
and the momentum of the parent in the laboratory frame)
is required to be less than 0.97.

Candidates for the decay $\eta^{\prime}\to\pi^+\pi^-\eta$ are reconstructed
using pairs of oppositely-charged pions and $\eta$ candidates. 
We require the reconstructed $\eta^{\prime}$ invariant mass to be in the range 
$920\le \mep \le 980$~${\rm MeV/c}^{2}$, which corresponds, 
approximately, to the range $[-10,+6]\sigma$ 
of the Gaussian approximation for the resolution function, 
after performing a kinematic fit 
constraining the reconstructed invariant mass 
of the daughter $\eta$ candidate 
to the nominal $\eta$ mass~\cite{PDG}. 
To identify charged pion candidates,
the ratios of PID likelihoods,
$R_{i/\pi}=L_{\pi}/(L_{\pi}+L_{i})$,
are used, where $L_{\pi}$  is the likelihood for the track according to pion hypothesis,
while $L_i$ is the likelihood according to kaon ($i=K$) or electron ($i=e$) hypotheses.
We require $R_{K/\pi}\le0.6$ and $R_{e/\pi}\le0.95$ for pion candidates.
According to MC studies, 
these requirements reject 
28\% of background signal candidates 
(which are primarily due to charged kaons and electrons), 
while the resulting efficiency loss is below 3\%. 
Charged pion tracks are required to originate from near the interaction point (IP)
by restricting their distance of closest approach along and perpendicular
to the beam collision axis to be less than 4.0~cm and 0.3~cm, respectively.
These selection criteria suppress beam-related backgrounds and reject poorly-reconstructed tracks. 
To reduce systematic uncertainties associated with track reconstruction efficiency, 
the transverse momenta of charged pions are required to be greater than 100~MeV/c. 

To identify $\bseep$ candidates we use
beam-energy constrained $B_s^0$ mass, $\mbc=\sqrt{E_{\rm beam}^2-p_{B_s}^2}$,
the energy difference, $\de=E_{B_s}-E_{\rm beam}$,
and the reconstructed invariant mass of the $\eta^\prime$,
where $E_{\rm beam}$, $p_{B_s}$ and $E_{B_s}$
are the beam energy,
the momentum magnitude
and
the reconstructed energy of $B_s^0$ candidate,
respectively.
All these quantities are evaluated in the $e^+e^-$ center-of-mass frame.
To improve the $\de$ resolution (by approximately 10\%),
each $\eta$ candidate is kinematically constrained to the nominal invariant mass of $\eta$,
the $\eta^{\prime}$ candidates are further constrained to the nominal invariant mass of $\eta^{\prime}$.
Signal candidates are required to satisfy selection criteria $\mbc>5.3$~${\rm GeV/c}^2$ and $-0.4 \le \de \le0.3$~GeV. 
In Gaussian approximation, the $\de$ resolution is, approximately, 40~MeV. 
The beam-energy-constrained $B_s^0$ mass resolution is 4~${\rm Mev/c^2}$. 
To improve the significance of the signal in case the data indicate its presence, 
we include the reconstructed invariant mass $\mep$ in the 3D ML fit 
used to statistically separate the signal from background.

\section{Background Suppression and Optimization for Discovery}

Hadronic continuum, {\it i.e.} production of light quark pairs
in the $e^+e^-$ annihilation [$e^+e^-\to q\bar{q}$ ($q=u,d,c,s$)],
is the primary source of background
in studies of charmless hadronic decays.
Because of large initial momenta of the light quarks,
continuum events exhibit a ``jet-like'' event shape,
while $\bsbsbara$ events are distributed isotropically.
We use modified Fox-Wolfram moments~\cite{ksfw}, 
used to describe the topology of the event, 
to discriminate between signal events and continuum background. 
A likelihood ratio ($\mathcal{LR}$) is calculated
using Fisher discriminant coefficients obtained 
in an optimization based on these moments. 
We suppress the background
using a discovery-optimized cut on $\mathcal{LR}$
obtained using Punzi's figure-of-merit~\cite{punzi}:

\begin{equation}
{\rm FOM} =\frac{\varepsilon(t) }{a/2+\sqrt{B(t)}},
\label{eq:FOM}
\end{equation}

\noindent where
$t$ is the cut on $\mathcal{LR}$,
$\varepsilon$ and $B$ are the overall signal reconstruction efficiency
and the number of background events expected in the signal region
for a given value of the cut on $\mathcal{LR}$, respectively.
The quantity $a$ is the desired significance 
(which we varied between 3 and 5) 
in the Gaussian approximation of Poisson statistics. 
To predict $B(t)$ we use sideband data with the signal region blinded 
and 
the scaling factor obtained from the background MC sample. 
We require signal candidates to satisfy the requirement $\mathcal{LR} \ge 0.95$, 
which corresponds to $B(0.95)=3$ and 52 background events in the 
signal and sideband regions of our fit variables, respectively. 

\begin{figure}
\small
 \begin{center}
    \subfigure{\includegraphics[width=.5\textwidth]{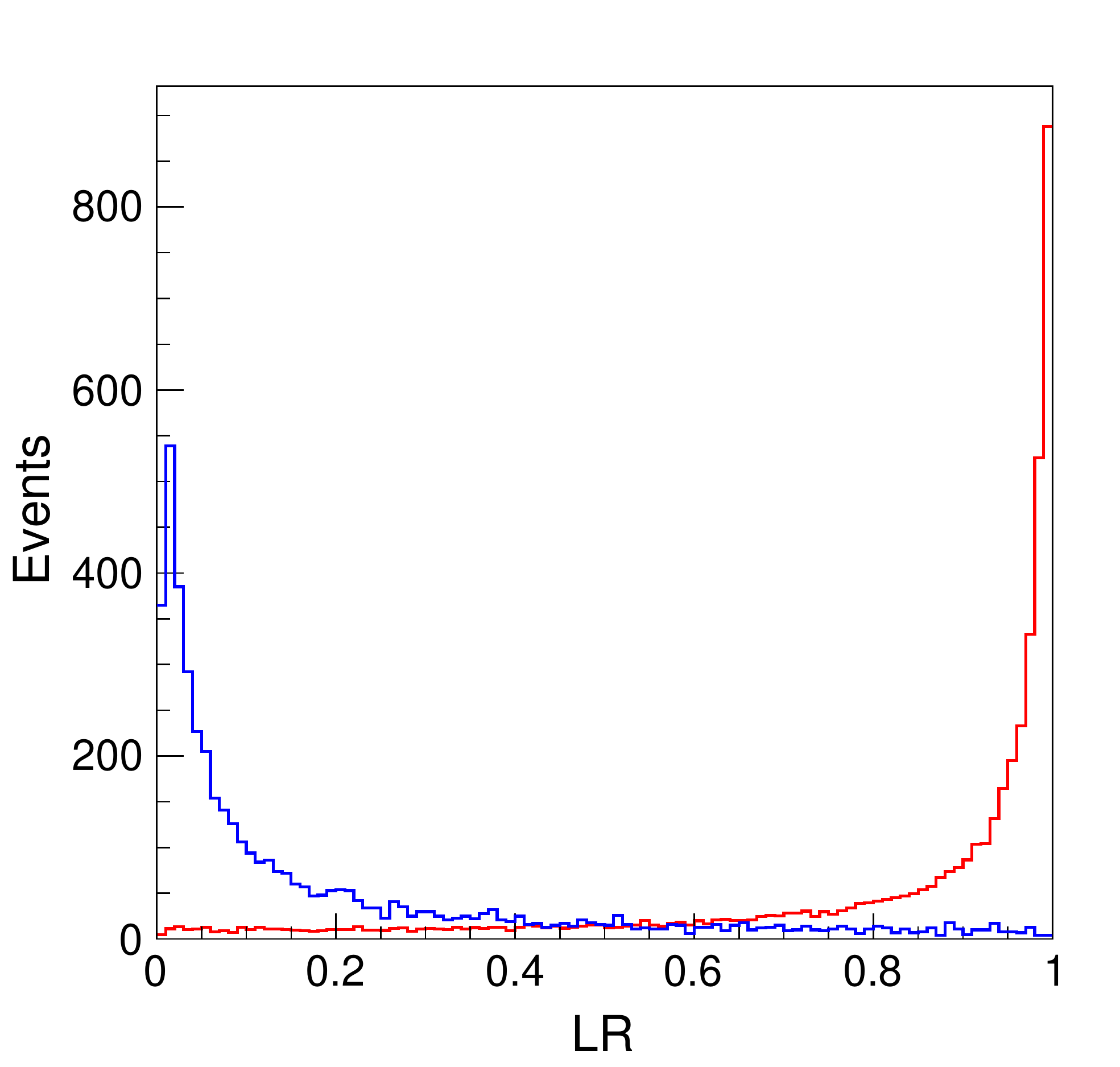}}
 \end{center}
\figcaption{Distributions of $\mathcal{LR}$
for signal (red) and background (blue) MC samples. 
Normalization is arbitrary.} 
\label{fig:lr}
\end{figure}

The background contains real $\eta^{\prime}$ mesons.
Such events exhibit a peak in the $\mep$ distribution,
however, they are distributed uniformly in $\mbc$ and $\de$.
The fraction of this peaking background 
is a free parameter in our ML fits.

\section{Candidate Multiplicity and Best Candidate Selection}

About 14\% of fully-reconstructed signal MC events contain multiple candidates
which are primarily (in 75\% of such events) due to misreconstructed $\eta$ mesons. 
In such events we use only the best candidate
with the smallest value of $\sum{\chi^2_{\eta}}+\chi^2_{\pi^+\pi^-}$,
where the values of $\chi^2_{\eta}$ are from the mass-constrained fit for the $\eta$ candidates
and $\chi^2_{\pi^+\pi^-}$ is from a vertex fit for the charged pion pair.
The overall reconstruction efficiency is estimated to be 10\%
including a 50\% relative efficiency loss
due to the discovery-optimized background suppression.

\section{Signal Extraction Procedure}

To extract the signal yield, we perform an unbinned extended maximum likelihood fit
to the three-dimensional (3D) distribution of $\mbc$, $\de$, and $\mep$.
The likelihood function is 

\begin{equation}
\mathcal{L}=\frac{e^{\sum_{j}n_j}}{N!}\prod_{i=1}^\textrm{N}\left(\sum_{j}n_{j}\mathcal{P}_{j}(M_{\rm bc}^i, \Delta E^i, M^i(\pi^+\pi^-\eta) )\right),
\label{eq:PDF}
\end{equation}

\noindent where the index $i$ is used for the events and $n_j$ are the fit parameters
describing the numbers of signal and background events.
Due to negligible correlations among fit variables for background and well-reconstructed signal events,
the probability densities are assumed to factorize as
$\mathcal{P}_{j}[M_{\rm bc}^i, \Delta E^i, M^i(\pi^+\pi^-\eta) ] =
\mathcal{P}_{j}(\mbc) \cdot \mathcal{P}_{j}(\de) \cdot \mathcal{P}_{j}[\mep]$.
The signal PDF is represented by the weighted sum of the 3D PDFs representing
possible $\bseep$ signal contributions from $\bsbsbara$ pairs,
where the weights are fixed according to previous measurements as described earlier:

\begin{equation}
\mathcal{P}_{sig}=\fbssbss \cdot \mathcal{P}_{B_{s}^{*0}\mybar{B}_s^{*0}} + \fbsbss \cdot \mathcal{P}_{B_{s}^{*0}\mybar{B}_s^0} + (1-\fbssbss-\fbsbss) \cdot \mathcal{P}_{B_{s}^0\mybar{B}_s^0}
\label{eqn:sigpdf}
\end{equation}

We use $B^0\to\eta^{\prime} K_S^0$ data recorded at the $\Upsilon(4S)$ resonance
to adjust the PDF shape parameters used to describe the signal.
We reconstruct $K_S^0$ candidates via secondary vertices 
associated with pairs of oppositely-charged pions~\cite{ks_reco} 
using a neural network (NN) technique~\cite{NN}. 
The following information is used in the NN: 
the momentum of $K_S^0$ candidate in the laboratory frame; 
the distance along the $z$ axis between 
the two track helices at the point of their closest approach; 
the flight length in the $x-y$ plane; 
the angle between the $K_S^0$ momentum 
and the vector joining the $K_S^0$ decay vertex to the IP; 
the angle between the pion momentum  and the laboratory-frame $K_S^0$ momentum in the $K_S^0$ rest frame; 
the distance-of-closest-approach in the $x-y$ plane between the IP and the two pion helices; 
and the pion hit information in the SVD and CDC. 
The selection efficiency is 87\% over the momentum range of interest. 
We also require that the $\pi^+\pi^-$ invariant mass 
be within 12~${\rm MeV/c^2}$ (about 3.5$\sigma$ in resolution) 
of the nominal $K_S^0$ mass~\cite{PDG}. 
We require $5.2\le \mbc \le 5.3$~${\rm GeV/c^2}$ for $B^0$ candidates. 
All other selection criteria applied to the $B^0$ candidates 
are the same as those used to select $B_s^0$ candidates. 

The presence of four photons in our final state gives rise to
a sizable misreconstruction probability for the signal.
We study partially misreconstructed signal events,
denoted Self Cross Feed (SCF) events, using signal MC sample.
A large correlation between $\mbc$ and $\de$ for such signal MC events 
(the Pearson correlation coefficient of 27\% for the region of largest same-sign correlations) 
is taken into account by describing the well-reconstructed part of the signal 
and SCF separately. 
SCF events comprise approximately 19\% of the reconstructed signal MC sample 
and are excluded from signal fit model and the efficiency estimate. 
No sizable correlations among fit variables 
have been identified for well-reconstructed signal MC events 
nor for background events.

\section{Fitting Models}

A sum of a Gaussian and a Crystal Ball~\cite{xbal} function is used to model the signal in each of the three fit variables.
For $M_{\rm bc}$ and $M(\pi^+\pi^-\eta)$ we use a sum with the same mean
but different widths, while for $\Delta E$ both mean and width are different. 
A different approach for the $\Delta E$ parametrization is necessary 
to provide a better description of its PDF which has a long asymmetric tail 
due to the additional particles used to evaluate this variable. 
We use a Crystal Ball function to describe the tails arising from energy leakage expected for photons in the calorimeter.
A Bukin function~\cite{bukin} and an asymmetric Gaussian are used to model the SCF contribution 
in $M_{\rm bc}$ and $\Delta E$, respectively. 
For $M(\pi^+\pi^-\eta)$, we use a sum of a Gaussian and 
a first order Chebyshev polynomial. 
The signal PDF shape parameters for $\mbc$ and $\de$
have been adjusted using the results obtained from the $\Upsilon(4S)$ data.

An ARGUS~\cite{argus} function is used to describe the background distribution in $M_{\rm bc}$, 
another first-order Chebyshev polynomial is used for $\Delta{E}$. 
To model the peaking part in $M(\pi^+\pi^-\eta)$ we use the signal PDF, 
because the peak is due to real $\eta^{\prime}$ mesons, 
while an additional first-order Chebyshev polynomial is used for non-peaking contribution. 

\section{Ensemble Tests}

\begin{figure}
  \small
  \begin{center}
    \subfigure{\includegraphics[width=0.5\textwidth]{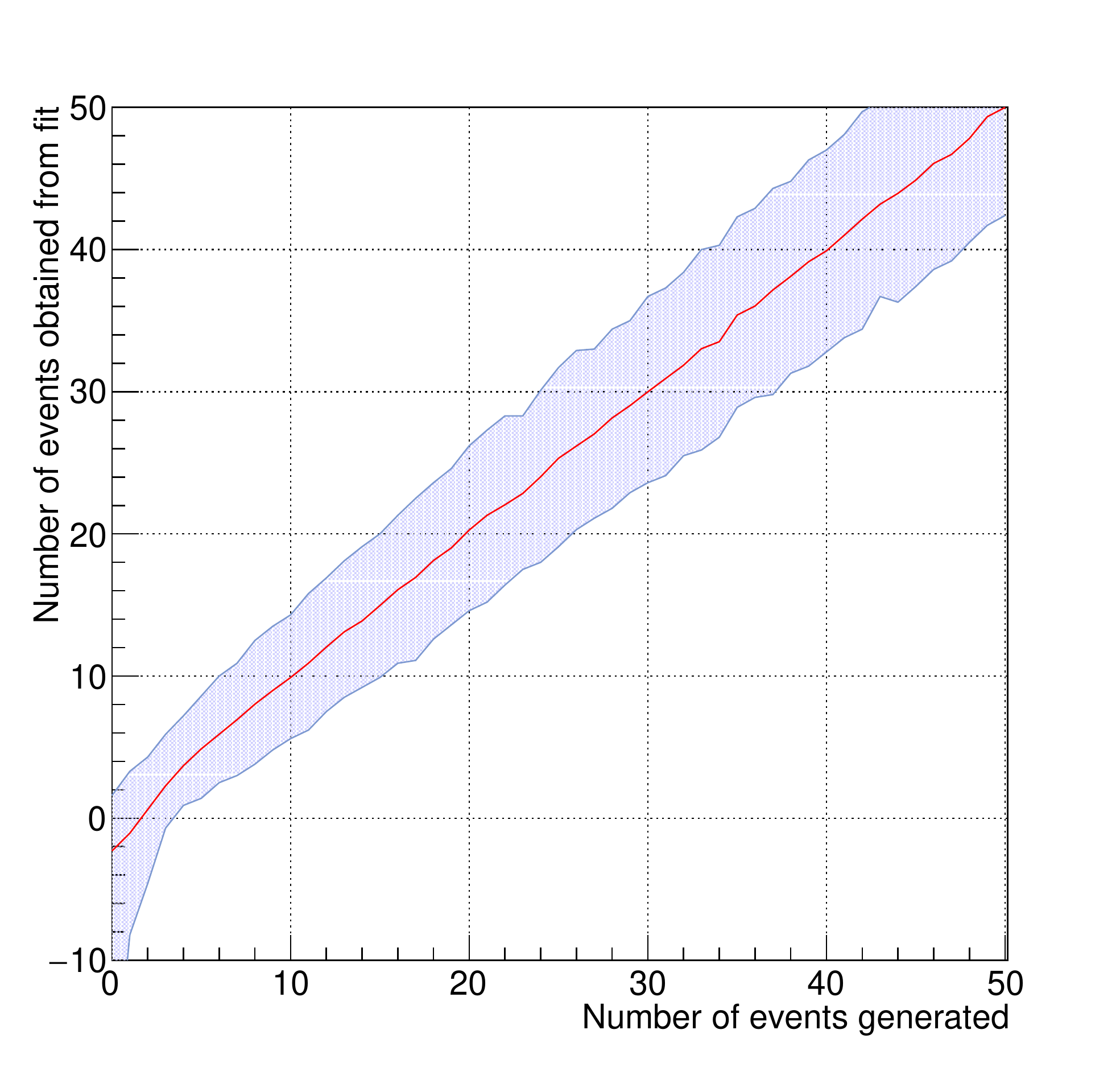}}
  \end{center}
  \figcaption{Classical 80\% confidence belt (shown by a blue band) obtained from pseudoexperiments.} 
  \label{cl_band}
\end{figure}

To test and validate our fitting model, ensemble tests are performed by generating MC pseudoexperiments. 
In these experiments we use PDFs obtained from simulation and the $\bksep$ data. 
The number of signal events is varied between 0 and 50 events, 
and 1000 pseudoexperiments are performed for each assumed number of signal events. 
An ML fit is performed for each sample generated in these experiments. 
For all values of assumed number of signal events 
the fit signal yield distribution peaks 
at the expected value, therefore exhibiting good linearity. 
We use the results of pseudoexperiments 
to construct classical confidence intervals (without ordering) 
using a procedure due to Neyman~\cite{frequentist_approach}. 
For each ensemble of pseudoexperiments 
the lower and upper ends of respective confidence interval 
represent the values of fit signal yields 
for which 10\% of the results lie below and above these values, 
respectively. 
These confidence intervals are then used to prepare 
a classical 80\% confidence belt~\cite{belt_method} 
shown in Fig.~\ref{cl_band}. 
We use this confidence belt 
to make a statistical interpretation 
of the results obtained from ML fit to data. 


\section{Results}

\begin{figure}
\small
\begin{center}
\subfigure{\includegraphics[width=1\textwidth]{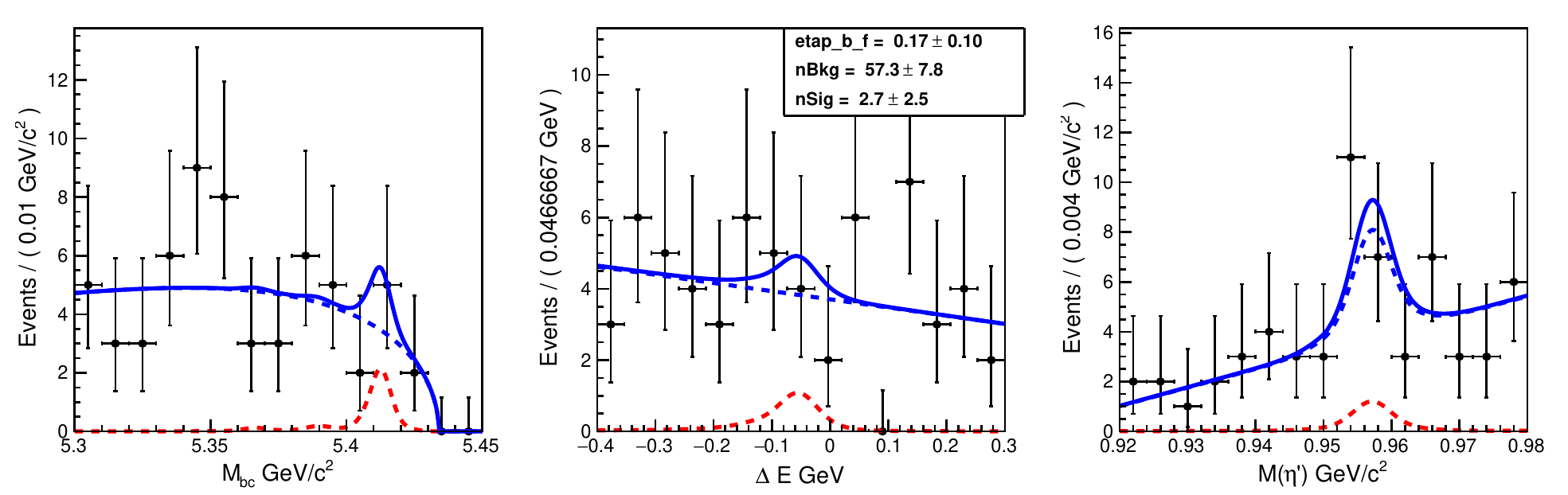}}
\subfigure{\includegraphics[width=1\textwidth]{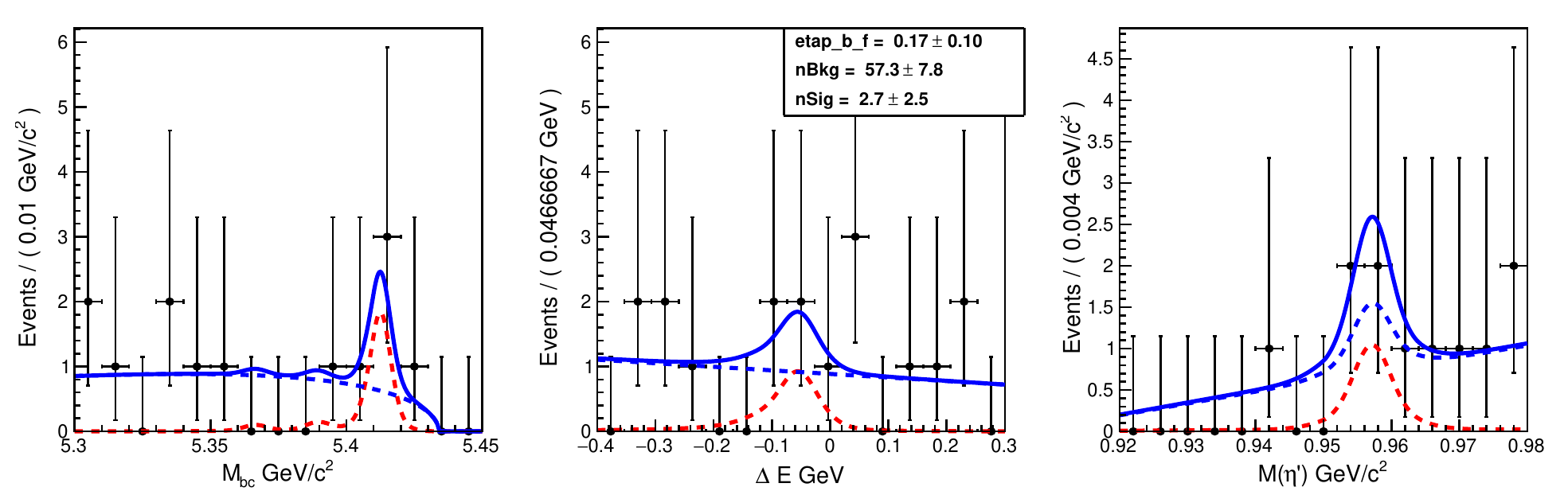}}
\end{center}
\figcaption{Full and signal-region projections of the 3D fit to the full $\Upsilon(5S)$ data sample. 
Signal and background PDFs are described in the text.}
\label{fit_data}
\end{figure}

We fit the 3D fit model described above to the data and
obtain $2.7 \pm 2.5$ signal and $57.3 \pm 7.8$ background events.
We show the signal-region projections of the fit to data in Fig.~\ref{fit_data}.
We observe no signal and estimate the 90\% confidence-level (CL) upper limit on the branching fraction
for the decay $\bseep$ using the frequentist approach~\cite{frequentist_approach} 
and the following formula:

\begin{equation}
\mathcal{B}(\bseep) < \frac{N_{\textrm{UL}}^{90\%}}{2 \cdot N_{\bsbsbara} \cdot \varepsilon \cdot \mathcal{B}_{\textrm{dp}}},
\label{eq_ul}
\end{equation}

\noindent where $N_{\bsbsbara}$ is the number of $\bsbsbara$ pairs in the full Belle data sample,
$\varepsilon$ is the overall reconstruction efficiency for the signal $B_s^0$ decay,
and $\mathcal{B}_{\textrm{dp}}$ 
is the product of the secondary branching fractions for all daughter particles in our final state.
Further, $N_{\textrm{UL}}^{90\%}$ is the expected signal yield
at 90\% CL which is the value representing the right side of the confidence belt at the central value for
signal yield, which is approximately 6 events. Using Eq.~(\ref{eq_ul}) we
obtain a 90\% CL upper limit on the branching fraction
of $\mathcal{B}(\bseep) < 7.1 \times 10^{-5}$.

\section{Systematics}

The relative systematic uncertainties for the quantities used in the upper limit estimate 
are summarized in Table~\ref{tab:lkr_sys}.  
The statistical uncertainty on the reconstruction efficiency
can be estimated as $\sqrt{\varepsilon\times (1-\varepsilon)/N}$,
where $N$ is the total number of generated signal MC events
and $\varepsilon$ is the reconstruction efficiency.
This uncertainty is estimated to be 0.1\%. 
We assign a 2.1\% systematic uncertainty per $\eta$ candidate~\cite{eta_syst}. 
Since we have two $\eta$ candidates in our decay, 
we assign a 4.2\% uncertainty for $\eta$ reconstruction.
The systematic uncertainty associated with
track reconstruction is 0.35\% per track~\cite{track_syst}. 
We therefore assign an uncertainty of 0.7\% for two tracks.
We assign a 15.3\% systematic uncertainty due to the discovery-optimized $\mathcal{LR}$ cut. 
This uncertainty reflects the relative change in the efficiency
when the cut is varied by 0.02 about nominal value of 0.95.
Combining all the sources of uncertainties, the total relative systematic uncertainty is 19\%.

\begin{table}[htb!]
\centering
\bgroup
\def\arraystretch{1.25}
\begin{tabular}{lc}
\hline \hline
Source     & Uncertainty (\%) \\
\hline \hline

Number of $\bsbsbara$ pairs            & 10.1 \\
Branching fraction of $\eta$           & 0.5  \\
Branching fraction of $\eta^{\prime}$  & 1.2  \\
MC statistics                          & 0.1  \\
$\eta$  reconstruction                 & 4.2  \\
Tracking                               & 0.7  \\
$\mathcal{LR}$ selection               & 15.3 \\
\hline
\hline
\end{tabular}
\egroup
\caption{Summary of systematic uncertainties in the $\bseep$ analysis.}
\label{tab:lkr_sys}
\end{table}

\section{Conclusions}

In summary, we have used the full data sample recorded 
by the Belle experiment at the $\Upsilon(5S)$ resonance 
to search for the rare decay $\bseep$. 
We observe no statistically significant signal and set a 90\% CL upper 
limit of $7.1 \times 10^{-5}$ on its branching fraction. 
Our result is 2 times larger than the most optimistic SM-based 
and QCD-enhanced theoretical prediction and, to date,  
is the only experimental information on $\bseep$.  
This decay will be probed further at the 
next-generation Belle~II experiment~\cite{belle2} 
at the SuperKEKB collider in Japan. 

\section{Acknowledgements}

We thank the KEKB group for the excellent operation of the
accelerator, the KEK cryogenics group for the efficient
operation of the solenoid, and the KEK computer group and
the National Institute of Informatics for valuable computing
and SINET3 network support. We acknowledge support from
the Ministry of Education, Culture, Sports, Science, and
Technology of Japan and the Japan Society for the Promotion
of Science; the Australian Research Council and the
Australian Department of Education, Science and Training;
the National Natural Science Foundation of China under
contract No.~10575109 and 10775142; the Department of
Science and Technology of India;
the BK21 program of the Ministry of Education of Korea,
the CHEP SRC program and Basic Research program
(grant No.~R01-2005-000-10089-0) of the Korea Science and
Engineering Foundation, and the Pure Basic Research Group
program of the Korea Research Foundation;
the Polish State Committee for Scientific Research;
the Ministry of Education and Science of the Russian
Federation and the Russian Federal Agency for Atomic Energy;
the Slovenian Research Agency;  the Swiss
National Science Foundation; the National Science Council
and the Ministry of Education of Taiwan; and the 
U.S.~Department of Energy.


\begin{thebibliography}{10}
\expandafter\ifx\csname natexlab\endcsname\relax\def\natexlab#1{#1}\fi
\expandafter\ifx\csname bibnamefont\endcsname\relax
  \def\bibnamefont#1{#1}\fi
\expandafter\ifx\csname bibfnamefont\endcsname\relax
  \def\bibfnamefont#1{#1}\fi
\expandafter\ifx\csname citenamefont\endcsname\relax
  \def\citenamefont#1{#1}\fi
\expandafter\ifx\csname url\endcsname\relax
  \def\url#1{\texttt{#1}}\fi
\expandafter\ifx\csname urlprefix\endcsname\relax\def\urlprefix{URL }\fi
\providecommand{\bibinfo}[2]{#2}
\providecommand{\eprint}[2][]{\url{#2}}

\bibitem[{\citenamefont{Artuso et~al.}(2008)}]{Buchalla:2008jp}
\bibinfo{author}{\bibfnamefont{M.}~\bibnamefont{Artuso}} \bibnamefont{et~al.},
  \bibinfo{journal}{Eur. Phys. J.} \textbf{\bibinfo{volume}{C57}},
  \bibinfo{pages}{309} (\bibinfo{year}{2008}), \eprint{0801.1833}.

\bibitem[{\citenamefont{Hsiao et~al.}(2016)\citenamefont{Hsiao, Chang, and
  He}}]{U3}
\bibinfo{author}{\bibfnamefont{Y.-K.} \bibnamefont{Hsiao}},
  \bibinfo{author}{\bibfnamefont{C.-F.} \bibnamefont{Chang}}, \bibnamefont{and}
  \bibinfo{author}{\bibfnamefont{X.-G.} \bibnamefont{He}},
  \bibinfo{journal}{Phys. Rev. {\rm \bf D}} \textbf{\bibinfo{volume}{93}},
  \bibinfo{pages}{114002} (\bibinfo{year}{2016}),
  \urlprefix\url{http://link.aps.org/doi/10.1103/PhysRevD.93.114002}.

\bibitem[{\citenamefont{{\it et al.}~(LHCb~Collaboration)}(2015)}]{LHCB}
\bibinfo{author}{\bibfnamefont{R.~A.} \bibnamefont{{\it et
  al.}~(LHCb~Collaboration)}} (\bibinfo{collaboration}{LHCb}),
  \bibinfo{journal}{Phys. Rev. Lett.} \textbf{\bibinfo{volume}{115}},
  \bibinfo{pages}{051801} (\bibinfo{year}{2015}), \eprint{1503.07483}.

\bibitem[{\citenamefont{Williamson and Zupan}(2006)}]{SCET}
\bibinfo{author}{\bibfnamefont{A.~R.} \bibnamefont{Williamson}}
  \bibnamefont{and} \bibinfo{author}{\bibfnamefont{J.}~\bibnamefont{Zupan}},
  \bibinfo{journal}{Phys. Rev. D} \textbf{\bibinfo{volume}{74}},
  \bibinfo{pages}{014003} (\bibinfo{year}{2006}).

\bibitem[{\citenamefont{Ali et~al.}(2007)\citenamefont{Ali, Kramer, Li, L\"u,
  Shen, Wang, and Wang}}]{pQCD}
\bibinfo{author}{\bibfnamefont{A.}~\bibnamefont{Ali}},
  \bibinfo{author}{\bibfnamefont{G.}~\bibnamefont{Kramer}},
  \bibinfo{author}{\bibfnamefont{Y.}~\bibnamefont{Li}},
  \bibinfo{author}{\bibfnamefont{C.-D.} \bibnamefont{L\"u}},
  \bibinfo{author}{\bibfnamefont{Y.-L.} \bibnamefont{Shen}},
  \bibinfo{author}{\bibfnamefont{W.}~\bibnamefont{Wang}}, \bibnamefont{and}
  \bibinfo{author}{\bibfnamefont{Y.-M.} \bibnamefont{Wang}},
  \bibinfo{journal}{Phys. Rev. D} \textbf{\bibinfo{volume}{76}},
  \bibinfo{pages}{074018} (\bibinfo{year}{2007}),
  \urlprefix\url{http://link.aps.org/doi/10.1103/PhysRevD.76.074018}.

\bibitem[{\citenamefont{Cheng and Chua}(2009)}]{QCDF}
\bibinfo{author}{\bibfnamefont{H.-Y.} \bibnamefont{Cheng}} \bibnamefont{and}
  \bibinfo{author}{\bibfnamefont{C.-K.} \bibnamefont{Chua}},
  \bibinfo{journal}{Phys. Rev. D} \textbf{\bibinfo{volume}{80}},
  \bibinfo{pages}{114026} (\bibinfo{year}{2009}),
  \urlprefix\url{http://link.aps.org/doi/10.1103/PhysRevD.80.114026}.

\bibitem[{\citenamefont{Cheng et~al.}(2015)\citenamefont{Cheng, Chiang, and
  Kuo}}]{SU3}
\bibinfo{author}{\bibfnamefont{H.-Y.} \bibnamefont{Cheng}},
  \bibinfo{author}{\bibfnamefont{C.-W.} \bibnamefont{Chiang}},
  \bibnamefont{and} \bibinfo{author}{\bibfnamefont{A.-L.} \bibnamefont{Kuo}},
  \bibinfo{journal}{Phys. Rev. D} \textbf{\bibinfo{volume}{91}},
  \bibinfo{pages}{014011} (\bibinfo{year}{2015}),
  \urlprefix\url{http://link.aps.org/doi/10.1103/PhysRevD.91.014011}.

\bibitem[{\citenamefont{Kurokawa and Kikutani}(2003)}]{KEKB}
\bibinfo{author}{\bibfnamefont{S.}~\bibnamefont{Kurokawa}} \bibnamefont{and}
  \bibinfo{author}{\bibfnamefont{E.}~\bibnamefont{Kikutani}},
  \bibinfo{journal}{Nuclear Instruments and Methods A}
  \textbf{\bibinfo{volume}{499}}, \bibinfo{pages}{1 } (\bibinfo{year}{2003}),
  ISSN \bibinfo{issn}{0168-9002},
  \urlprefix\url{http://www.sciencedirect.com/science/article/pii/S0168900202017710}.

\bibitem[{\citenamefont{Lange}(2001)}]{EvtGen}
\bibinfo{author}{\bibfnamefont{D.~J.} \bibnamefont{Lange}},
  \bibinfo{journal}{Nuclear Instruments and Methods A}
  \textbf{\bibinfo{volume}{462}}, \bibinfo{pages}{152 } (\bibinfo{year}{2001}),
  ISSN \bibinfo{issn}{0168-9002},
  \urlprefix\url{http://www.sciencedirect.com/science/article/pii/S0168900201000894}.

\bibitem[{\citenamefont{Agostinelli et~al.}(2003)}]{GEANT4}
\bibinfo{author}{\bibfnamefont{S.}~\bibnamefont{Agostinelli}}
  \bibnamefont{et~al.}, \bibinfo{journal}{Nuclear Instruments and Methods A}
  \textbf{\bibinfo{volume}{506}}, \bibinfo{pages}{250 } (\bibinfo{year}{2003}),
  ISSN \bibinfo{issn}{0168-9002},
  \urlprefix\url{http://www.sciencedirect.com/science/article/pii/S0168900203013688}.

\end{thebibliography}


\begin{thebibliography}{99}

\bibitem{belleiiphysicsbook}
E.~Kou, P.~Urquijo, W.~Altmannshofer {\it et al.} 
(Belle~II Collaboration), 
Prog Theor Exp Phys (2019), 
arXiv:1808.10567 [hep-ex]. 

\bibitem{bf1}
Y.-K.~Hsiao, C.-F.~Chang, and X.-G.~He,
Phys. Rev. D {\bf 93}, 114002 (2016). 

\bibitem{bsepep}
R. Aaij {\it et al.} (LHCb Collaboration),
Phys. Rev. Lett.{\bf 115}, 051801 (2015).

\bibitem{Belle}
A.~Abashian {\it et al.} (Belle Collaboration),
Nucl. Instr. and Meth. A {\bf 479}, 117 (2002).

\bibitem{KEKB}
S.~Kurokawa and E.~Kikutani, Nucl. Instr. and. Meth. A499, 1 (2003),
and other papers included in this volume. 

\bibitem{frac}
S. Esen {\it et al.} (Belle Collaboration),
Phys. Rev. D {\bf 87}, 031101(R) (2013). 

\bibitem{nbsbsb}
C. Oswald {\it et al.} (Belle Collaboration), 
Phys. Rev. D {\bf 92}, 072013 (2015). 


\bibitem{bf2}
A.~R.~Williamson and J.~Zupan,
Phys. Rev. D {\bf 74}, 014003 (2006). 

\bibitem{bf3}
A.~Ali, G.~Kramer, Y.~Li, C.-D.~L${\rm \ddot{u}}$ {\it et al.}, 
Phys. Rev. D  {\bf 76}, 074018 (2007). 

\bibitem{bf4}
H.-Y.~Cheng and C.-K.~Chua,
Phys. Rev. D {\bf 80}, 114026 (2009). 

\bibitem{bf5}
H.-Y.~Cheng, C.-W.~Chiang, and A.-L.~Kuo,
Phys. Rev. D {\bf 91}, 014011 (2015). 

\bibitem{PDG}
P.A.~Zyla {\it et al.} (Particle Data Group), Prog. Theor. Exp. Phys. {\bf 2020}, 083C01 (2020).

\bibitem{ksfw}
The Fox-Wolfram moments were introduced in G.~C.~Fox 
and S.~Wolfram, Phys. Rev. Lett. {\bf 41}, 1581 (1978). The
Fisher discriminant used by Belle, based on modified Fox-Wolfram moments, 
is described in K.~Abe {\it et al.}
(Belle Collaboration), 
Phys. Rev. Lett. {\bf 87}, 101801 (2001)
and 
K.~Abe {\it et al.} (Belle Collabboration.), Phys. Lett. B {\bf 511}, 151 (2001). 

\bibitem{punzi}
G.~Punzi, 
eConf C {\bf 030908} (2003), 
{\it Proceedings of PHYSTAT2003: 
Statistical Problems in Particle Physics, Astrophysics and Cosmology},  
arXiv:physics/0308063 [physics.data-an]. 

\bibitem{ks_reco}
K.-F.~Chen {\it et al.} (Belle Collaboration), Phys. Rev. D {\bf 72}, 012004 (2005). 

\bibitem{NN}
M.~Feindt and U.~Kerzel, 
{\it The NeuroBayes neural network package}, 
Nucl. Instrum. Methods Phys. Res., 
Sect.~A~{\bf 559}, 190 (2006). 

\bibitem{xbal}
M.~Oreglia, {\it A Study of the Reactions $\psi^\prime \to \gamma\gamma\psi$.} PhD thesis, SLAC, 1980. \\
T.~Skwarnicki, {\it A study of the radiative CASCADE transitions between the Upsilon-Prime and Upsilon resonances.} PhD thesis, Cracow, INP, 1986. 

\bibitem{bukin}
A.D.~Bukin, 
{\it Fitting function for asymmetric peaks}, 
arXiv:0711.4449 [physics.data-an] (2007). 

\bibitem{argus}
H.~Albrecht {\it et al.} (ARGUS Collaboration), Phys. Lett. {\bf B} 241, 278 (1990). 

\bibitem{belt_method}
A.~Stuart and J.K.~Ord, {\it Classical Inference and Relationship},
5th ed., 
Kendall's Advanced Theory of Statistics, 
Vol.~2 (Oxford University Press, New York, 1991); 
see also earlier editions by Kendall and Stuart. \\
W.T.~Eadie, D.~Drijard, F.E.~James, M.~Roos, and B.~Sadoulet, 
{\it Statistical Methods in Experimental Physics}, 
(NorthHolland, Amsterdam, 1971). 

\bibitem{frequentist_approach}
J.~Neyman, Phil. Trans. Roy. Soc. Lond. {\bf A236}, 767, 333 (1937); 
Reprinted in 
{\it A Selection of Early Statistical Papers of J. Neyman}, 
(University of California Press, Berkeley, 1967).

\bibitem{eta_syst}
J.~Schumann {\it et al.} (Belle Collaboration),
Phys. Rev. Lett. {\bf 97}, 061802 (2006). 

\bibitem{track_syst}
S.~Ryu {\it et al.} (Belle Collaboration), 
Phys. Rev. {\bf D} 89, 072009 (2014). 


\bibitem{belle2}
T. Abe {\it et al.} (Belle~II Collaboration), arXiv:{\bf 1011.0352} [physics.ins-det] (2010).

\end{thebibliography}

%

\end{document}